# Quantum oscillations from a two-dimensional electron gas at a Mott/band insulator interface

Pouya Moetakef[1], Daniel G. Ouellette[2], James R. Williams[3], S. James Allen[2], Leon Balents[2], David Goldhaber-Gordon[3], and Susanne Stemmer[1,a)]

[1]Materials Department, University of California, Santa Barbara, California, 93106-5050, USA

[2]Department of Physics, University of California, Santa Barbara, California, 93106-9530, USA

[3]Department of Physics, Stanford University, Stanford, California, 94305-4045, USA

**Abstract**

We report on the magnetotransport properties of a prototype Mott insulator/band insulator perovskite heterojunction in magnetic fields up to 31 T and at temperatures between 360 mK and 10 K. Shubnikov-de Haas oscillations in the magnetoresistance are observed. The oscillations are two-dimensional in nature and are interpreted as arising from either a single, spin-split subband or two subbands. In either case, the electron system that gives rise to the oscillations represents only a fraction of the electrons in the space charge layer at the interface. The temperature dependence of the oscillations are used to extract an effective mass of ~ 1 $m_e$ for the subband(s). The results are discussed in the context of the $t_{2g}$-states that form the bottom of the conduction band of $SrTiO_3$.

[a] Electronic mail: stemmer@mrl.ucsb.edu

The discovery of mobile charge carriers at oxide interfaces [1] has opened up new scientific opportunities [2,3], such as two-dimensional electron gases in materials where transport occurs in *d*-electron bands, which are subject to strong electron-electron interactions. Shubnikov-de Haas oscillations of the longitudinal resistance in a quantizing magnetic field elucidate the properties of the carriers and establish the two-dimensional character of electron gases [4]. The Shubnikov-de Haas oscillation period for a two-dimensional system depends only on the component of the magnetic field (*B*) that is perpendicular to the plane. Several studies have reported on Shubnikov-de Haas oscillations from $LaAlO_3/SrTiO_3$ interfaces [5-7]. However, open questions remain, including regarding the dimensionality of the electron system and the complete interpretation of the oscillations. In one study oscillations were found to depend only on the perpendicular field, pointing to a two-dimensional electron system [5], while another study found the angle dependence to indicate departure from two-dimensionality [7]. Complexity arises from the manifold of states that exist at the conduction band minimum of $SrTiO_3$ [8,9], and the occupation of multiple subbands at typical electron densities at oxide interfaces. Furthermore, mobile electron densities for $LaAlO_3/SrTiO_3$ interfaces are approximately an order of magnitude lower (~ $10^{13}$ cm$^{-2}$) than the carrier density needed to compensate the polar discontinuity at the interface [3]. As a result, departure from strictly two-dimensional behavior may occur: the electrostatic confining potential is reduced at low electron densities [10]. The high dielectric constant of $SrTiO_3$ causes carriers to spread over a significant distance [10,11] and a small energy separation between subbands, possibly smaller than the Landau level splitting at accessible *B*-fields, adds complexity to the oscillations in tilted *B*-fields.

Further understanding of the electron gas formed at the interface between insulating oxide materials can be obtained by examining other interfaces expected to produce mobile electrons. For example, GdTiO$_3$/SrTiO$_3$ interfaces share similarities with LaAlO$_3$/SrTiO$_3$ interfaces, including an interfacial polar discontinuity. Both materials are insulators in their bulk, stoichiometric forms (GdTiO$_3$ is a Mott insulator with a 3$d$-3$d$ band gap of ~ 0.7 eV [12]). Moreover, these interfaces show mobile carrier densities, which reside on the SrTiO$_3$ side, of ~3×10$^{14}$ cm$^{-2}$, an order of magnitude higher than for LaAlO$_3$/SrTiO$_3$, and closely corresponding to the density needed to compensate the polar discontinuity [13]. The high carrier density results in a large confining potential, which reduces the field-dependent dielectric constant of SrTiO$_3$ [14], thus producing much tighter spatial confinement than in LaAlO$_3$/SrTiO$_3$. In this Letter, we demonstrate two-dimensional quantum oscillations from the electron gas at the GdTiO$_3$/SrTiO$_3$ interface. From the temperature dependence of these oscillations, an electron mass of ~ 1 $m_e$ is extracted, which allows for the identification of the specific $t_{2g}$ orbital, $d_{xy}$, as the source of the quantum oscillations.

GdTiO$_3$ and SrTiO$_3$ were grown on (001) (LaAlO$_3$)$_{0.3}$(Sr$_2$AlTaO$_6$)$_{0.7}$ (LSAT) single crystals by molecular beam epitaxy (MBE) [13,15,16]. The sample discussed in this Letter consisted of 80-nm-thick SrTiO$_3$ followed by ~ 8 nm (20 pseudocubic unit cells) of GdTiO$_3$. Atomic resolution scanning transmission electron microscopy (STEM) showed that interfaces were atomically abrupt [13,15]. Each GdTiO$_3$/SrTiO$_3$ interface introduced ~3×10$^{14}$ cm$^{-2}$ carrier per interface, independent of the growth sequence and layer thicknesses [13]. Magnetotransport measurements were performed in van der Pauw geometry on 10×10 mm$^2$ samples. Ohmic contacts were 50-nm-Ti/300-nm-Au. Sheet resistance and Hall measurements at low $B$ were carried out using a Physical Properties Measurement System (Quantum Design PPMS). The sheet carrier density was ~ 3×10$^{14}$ cm$^{-2}$ and n-type [17]. The Hall mobility at 2 K ($\mu_{tr}$) was 322 cm$^2$/Vs. Measurements of the Hall resistance, which was linear in $B$ up to $B$ = 31 T, and of the longitudinal resistance from 0 to 31 T where performed at the National High Magnetic Field Laboratory (NHMFL) using a $^3$He cryostat at temperatures between 360 mK to 10 K and a standard lock-in technique. Samples were mounted on a rotator probe for rotation to angles between 0° to 90°, where 0° indicates that $B$ is perpendicular to the interface. Measurements were carried using a 2 T/min sweep rate. For $B$ greater than 16 T, Shubnikov-de Haas oscillations were detected in measurements of the longitudinal resistance as a function of $B$ [17]. Another similarly prepared sample was investigated and showed similar results [17]. The non-oscillating background was subtracted using multiple polynomial fits to obtain the oscillating component of the resistance, $\Delta R_{xx}$.

The dimensionality of the electron system that gives rise to the Shubnikov-de Haas oscillations was probed by measuring the angular dependence of the oscillations. Figure 1(a) shows the oscillations measured at 360 mK as a function of $(B\cos\theta)^{-1}$, for different tilt angles $\theta$ of the $B$-field. Simpler characteristics compared to oscillations reported for LaAlO$_3$/SrTiO$_3$ interfaces [5-7] are observed. At $\theta = 0°$, oscillations are periodic in $1/B$ [red trace in Fig. 1(a)]. Maxima (and minima) appear at the same $(B\cos\theta)^{-1}$ value for all angles to ~ 40°, confirming the two-dimensionality of the electron system. The extracted oscillation frequency (primary peak, labeled $a$), obtained from a Fourier transform of $\Delta R_{xx}$ [inset of Fig. 1(b)] as a function of $1/B$, is shown in Fig. 1(b) for different $B$-field tilt angles. The product of the frequency and $\cos\theta$ is independent of $\theta$ ($a\cos\theta \sim 1171$ T), as expected for a two-dimensional system.

To further analyze the electron system that gives rise to the observed oscillations, Fig. 2 shows $\Delta R_{xx}$ at 360 mK as a function of $1/B$ ($B$ is normal to the interface). At low magnetic fields (< 22 T, as indicated by the upper right red box of Fig. 2), two dominant frequencies are present and can be related to the weak minima that appear between the strong minima for low $B$-fields, and maxima that alternate in height. Fourier transforms of this field range (upper inset of Fig. 2), confirm this, exhibiting two narrow peaks, at 637 T and 1170 T. At $B > 25$ T, the difference between the minima and maxima are less obvious (see also [17]) and the Fourier transform is dominated by a single peak at 1170 T (lower inset in Fig. 2). The presence of two narrow peaks, at 637 T ($f_b$) and 1170 T ($f_a$), in the Fourier transform, which are almost harmonically related, can be interpreted as being due to single, spin-split subband, giving rise to an oscillation frequency of $f_a = n_a h/e$, or due to two subbands, each having an oscillation frequency of $f_i = n_i h/2e$, where $e$ is the electron charge, $h$ is Planck's constant and $n_i$ ($i = a, b$) denotes the carrier density in the respective subband. The occupation of multiple subbands in this system is expected (see discussion below). Fits using a standard two-subband expression [18,19] for a two-dimensional electron gas with no spin-splitting resulted in a reasonable description of the experimentally observed frequencies, and the change in the sequence of alternating strong/weak oscillations [17]. However, a mismatch in the frequencies between experiment and simulations occurred at higher as well as lower $B$-fields. The subband splitting can be estimated to be 60 meV from the two frequencies as $f_a - f_b = (E_a - E_b)(m^*/e\hbar)$, where $m^*$ is the effective mass (determined below), and $(E_a - E_b)$ the energy difference between the subbands. The spacing would be consistent with the predicted energy spacing of very high lying $d_{xy}$ derived subbands [11,20]. Although it is not impossible in this system that an upper subband has an electron density just near half of the lower subband [10], the fact that the two frequencies are almost harmonically related, and that the two subbands have almost identical masses (see below), appears to favor more strongly an interpretation that the oscillations arise from a single, spin-split subband.

In the spin-splitting scenario, a second peak in the Fourier transform arises at low fields, because the spin-split minima are less deep than those associated with Landau level splitting, due to a smaller energy gap. Spin-related effects are expected in SrTiO$_3$ [21]: because of the heavy mass of electrons, the Landau level splitting, $\hbar\omega_c$, becomes comparable to the Zeeman energy, $E_z = \mu_B g^* B$, where $\hbar$ is the reduced Planck's constant, $\omega_c$ is the cyclotron frequency, $\mu_B$ is the Bohr magnetron and $g^*$ is the effective Landé-factor. There is no theory for the detailed shape of Shubnikov-de Haas oscillations of a two-dimensional, spin-split subband coexisting with other, lower-mobility subbands, and, at present, we cannot distinguish between the two scenarios.

For the scenario of a single spin-split subband, the carrier density in the subband is about $2.8 \times 10^{13}$ cm$^{-2}$. For the scenario of two subbands the total carrier density is $n_a + n_b = 3.1 \times 10^{13}$ cm$^{-2}$ + $5.7 \times 10^{13}$ cm$^{-2}$ = $8.7 \times 10^{13}$ cm$^{-2}$. These carrier densities represent a fraction (10% or 30%, respectively) of the carrier density measured from the Hall. Previous studies of LaAlO$_3$/SrTiO$_3$ interfaces [5] and delta-doped SrTiO$_3$ films [21-23] also reported carrier densities from Shubnikov-de Haas oscillations that were only a small fraction of the Hall density. The most likely explanation is that a large number of carriers reside in subbands that have a heavy mass or experience strong scattering, so that the Shubnikov-de Haas oscillations of these subbands are suppressed. The states at the bottom of the conduction band of SrTiO$_3$ are derived from the three (nearly) degenerate $t_{2g}$ d-orbitals of Ti ($d_{xy}$, $d_{xz}$, $d_{yz}$). The degeneracy is lifted by the confining electrostatic potential. The most tightly bound subbands have $d_{xy}$ character, with a low isotropic effective mass (~ 1 $m_e$) *in-plane* [11]. At these electron densities, the higher-lying $d_{xz}$ and $d_{yz}$ derived bands are also occupied [24]. These bands have an anisotropic mass in-plane, a low mass out-of-plane, and these electrons are expected to be less confined. Theoretical calculations confirm this basic picture [8,11,20,25,26].

The type of orbital from which the subband(s) giving rise to the oscillations is (are) derived can be identified by determining the effective electron mass, $m^*$, extracted from the temperature dependence of the oscillations (Fig. 3). Figure 3(b) shows the temperature dependence of the amplitudes of the Fourier transform peaks, which were fit to the following equation:

$$\frac{A(T)}{T} = \frac{A_0}{\sinh\left(\frac{2\pi^2 k_b T}{\hbar \omega_c}\right)}, \qquad (1)$$

where $A(T)$ is the amplitude of the peak at temperature $T$, and $k_b$ is the Boltzmann constant. The fits yielded $\omega_c$, which was used to obtain $m^*$, according to $\omega_c = eB/m^*$, which were 1.01 $m_e$ (peak $a$) and 1.08 $m_e$ (peak $b$), respectively. This low effective mass points to the oscillations being due to the more strongly confined $d_{xy}$-derived subbands, which also explains their strongly two-dimensional character. The cyclotron averaged mass of the $d_{xz}$ and $d_{yz}$ derived bands is at least $3m_e$ [9,10,20], in disagreement with our extracted value. Conversely, the results show that to obtain two-dimensional quantum oscillations from electron gases in SrTiO$_3$, the quality of the samples must be sufficient to allow for oscillations from the more tightly confined $d_{xy}$-derived subband(s), which are expected to suffer more strongly from interface scattering than the less confined $d_{xz}$ and $d_{yz}$ derived bands. The Fermi circle areas, derived from the periodicity in $1/B$, with the areas predicted by theoretical models [10,11,20], are in rough agreement with these predictions, if we identify the oscillations with the lowest $d_{xy}$ derived subbands and ignore possible complications due to coupling with $d_{xz}$ and $d_{yz}$ derived bands. Magnetic breakdown at high fields will decouple the simple orbits associated with the $d_{xy}$ bands from the orbits associated with the $d_{xz}$ and $d_{yz}$ [8,27].

To determine the quantum scattering time, the decay of the amplitudes was fit according to [18]:

$$\Delta R_{xx} = 4R_0 \exp\left(-2\pi^2 k_b T_D / \hbar\omega_c\right) \frac{2\pi^2 k_b T / \hbar\omega_c}{\sinh\left(2\pi^2 k_b T / \hbar\omega_c\right)}, \qquad (2)$$

where $T_D$ is the Dingle temperature. For the spin-split subband scenario, the Dingle temperature is 7 K, and the quantum scattering time, $\tau_q = \frac{\hbar}{2\pi k_b T_D}$, $1.7\times10^{-13}$ s, corresponding to a quantum mobility of 295 cm$^2$V$^{-1}$s$^{-1}$. Similar values were obtained for the two subband case [17]. The quantum scattering sets a lower limit on the transport scattering time in that subband [4,28]. The transport scattering time, $\tau_{tr}$, inferred from the Hall mobility, $\mu_{tr} = e\tau_{tr}/m^*$, was about $1.8\times10^{-13}$ s, and contains contributions from the carriers that are seen in the Hall measurements but do not give rise to the oscillations [17]. Future investigations should be directed towards comparisons of the Fermi surface area and masses with theory of the electronic structure of two-dimensional electron gases in SrTiO$_3$.

The authors thank Guru Khalsa, Allan MacDonald, and Mike Manfra for many useful discussions, Jonathan Billings and Alexey Suslov for assistance with the experiments at NHMFL and Tyler Cain for growing the sample. P.M. was supported by the U.S. National Science

Foundation (Grant No. DMR-1006640) and D. G. O by the MRSEC Program of the National Science Foundation (Award No. DMR 1121053). J. R. W., S. J. A., L.B., D. G-G. and S.S. acknowledge support by a MURI program of the Army Research Office (Grant No. W911-NF-09-1-0398. A portion of this work was performed at the National High Magnetic Field Laboratory, which is supported by National Science Foundation (Cooperative Agreement No. DMR-0654118), the State of Florida, and the U.S. Department of Energy.

**Figure Captions**

**Figure 1 (color online):** (a) Shubnikov de Haas oscillations measured at 360 mK as a function of $(B\cos\theta)^{-1}$ for different tilt angles $\theta$ of $B$ with respect to the interface normal. (b) Frequency of peak 'a' (open circles) in the Fourier transform (see inset) and its frequency multiplied with $\cos\theta$ (solid symbols) as a function of tilt angle. The dashed line is a linear fit. The inset shows a Fourier transform of $\Delta R_{xx}$ as a function of $1/B$ at $\theta = 0°$.

**Figure 2 (color online):** Shubnikov-de Haas oscillations ($\Delta R_{xx}$ as a function of $1/B$) at 360 mK and $\theta = 0°$. The solid and dashed arrows indicate strong and weak oscillations, respectively. The insets show Fourier transforms of the regions indicated by the red boxes. For small $B$, two frequencies are evident at 637 T and 1170 T. At higher values of $B$, a single frequency is measured, at a value of 1170 T.

**Figure 3 (color online):** (a) Temperature dependent Shubnikov-de Haas oscillations ($\Delta R_{xx}$ as a function of $1/B$) and $\theta = 0°$ at temperatures between 360 mK and 10 K. (b) Fourier transform of the data. The inset shows $A(T)/T$ of peak $a$ as a function of temperature and a fit according Eq. (1).

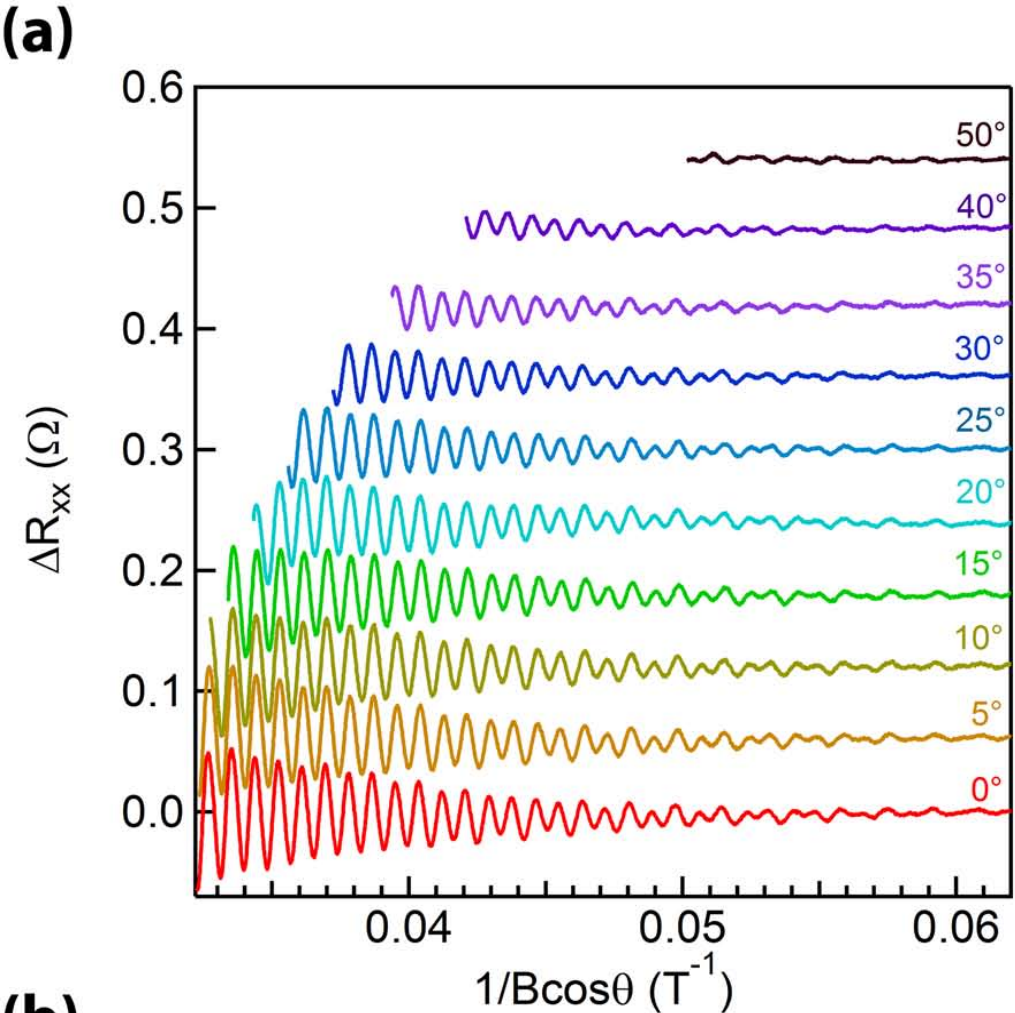

(a)

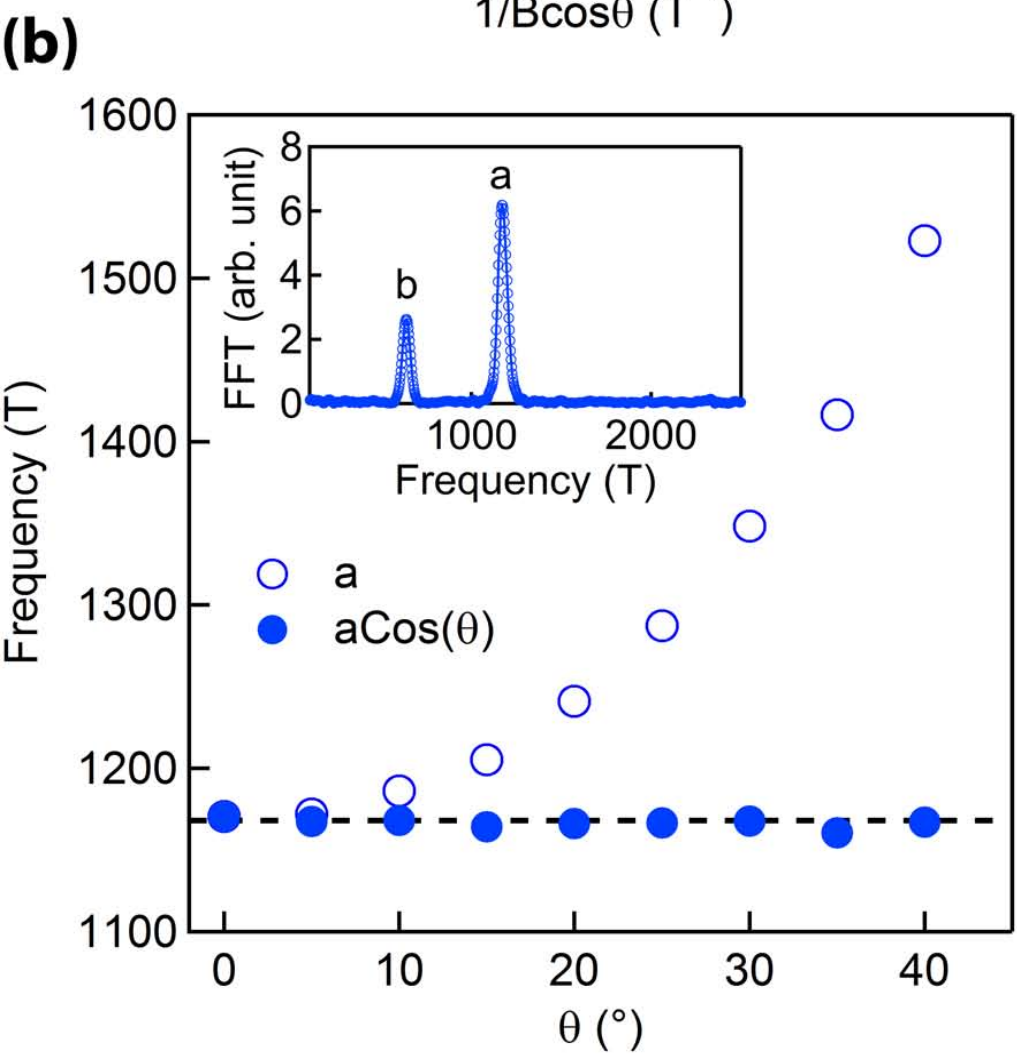

(b)

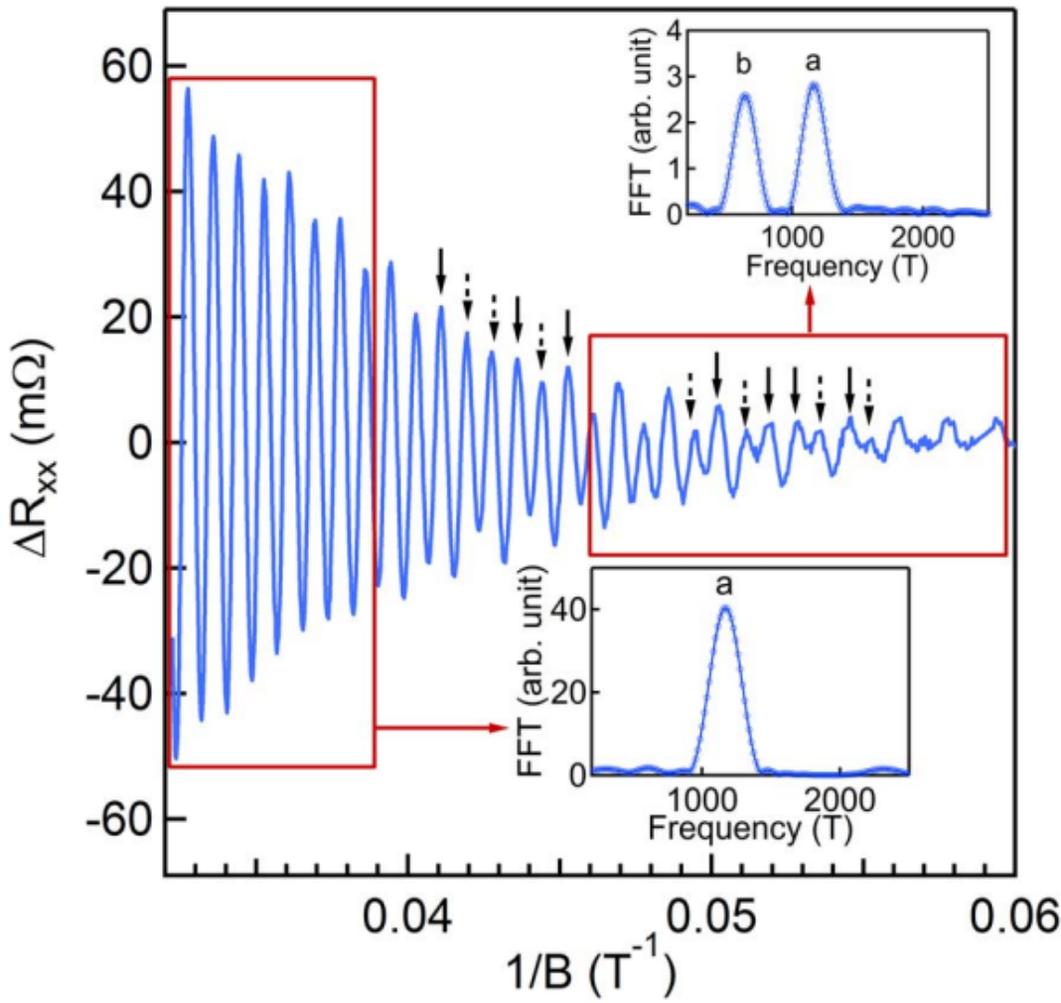

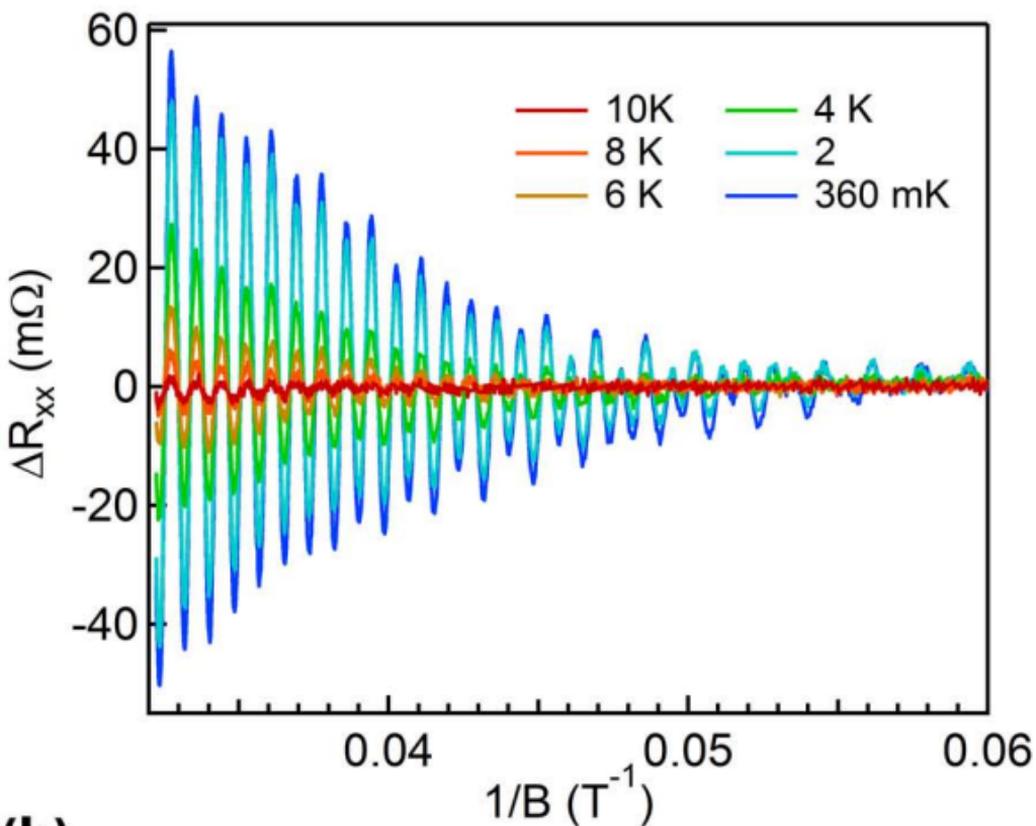
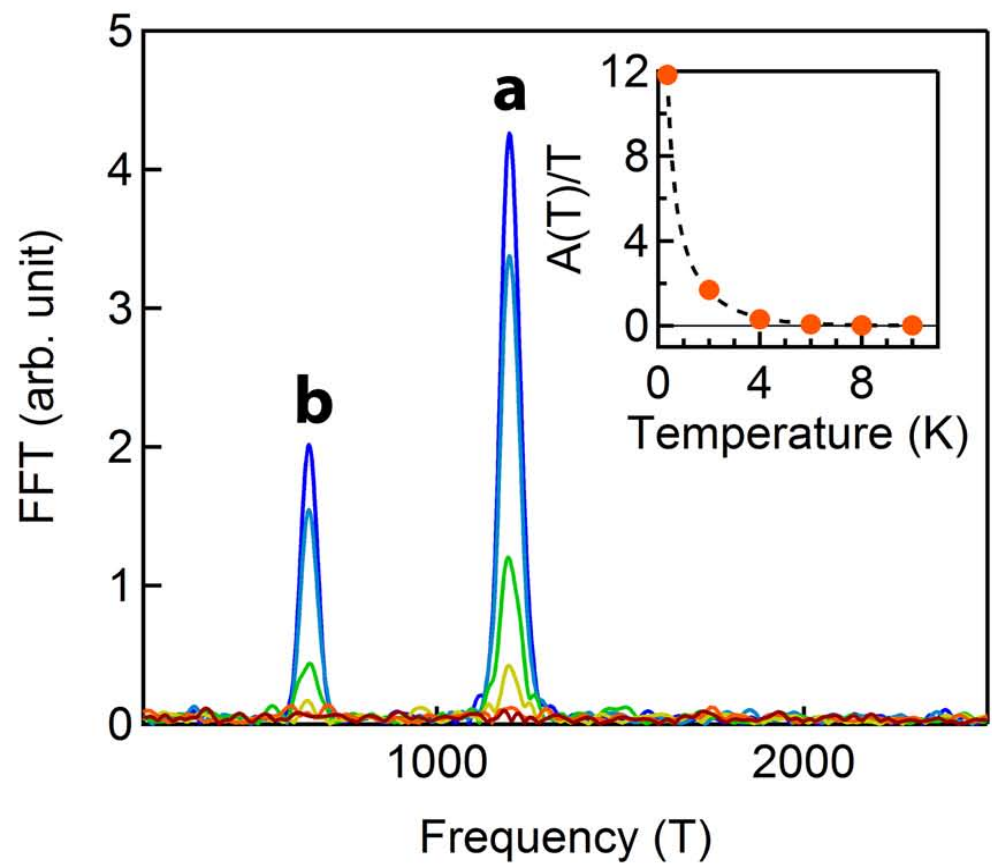